\documentstyle[11pt,epsf]{article}

\newcommand{\beq}{\begin{equation}} \newcommand{\eeq}{\end{equation}}
\newcommand{\bea}{\begin{eqnarray}} \newcommand{\eea}{\end{eqnarray}}

  \newcommand
{\Romannumeral}[1]{\uppercase\expandafter{\romannumeral#1}}

\newcommand{\be}{\begin{enumerate}} \newcommand{\ee}{\end{enumerate}}
\newcommand{\bi}{\begin{itemize}} \newcommand{\ei}{\end{itemize}}
\newcommand{\ba}{\begin{array}} \newcommand{\ea}{\end{array}}
\newcommand{\bc}{\begin{center}} \newcommand{\ec}{\end{center}}
\newcommand{\bt}{\begin{tabular}} \newcommand{\et}{\end{tabular}}

\def\lsim{\mathrel{\rlap{\lower4pt\hbox{\hskip1pt$\sim$}}
    \raise1pt\hbox{$<$}}}           
\def\gsim{\mathrel{\rlap{\lower4pt\hbox{\hskip1pt$\sim$}}
    \raise1pt\hbox{$>$}}}           

\newcommand{\half}{\textstyle {1\over2} \displaystyle}    
\newcommand{\third}{\textstyle {1\over3} \displaystyle}   
\newcommand{\Dslash}{{\hbox{D}\kern-0.6em\raise0.15ex\hbox{/}}} 

\begin{document}

\setlength{\oddsidemargin}{0cm} \setlength{\baselineskip}{7mm}

\input epsf

\begin{normalsize}\begin{flushright}
DAMTP-2005-58 \\
June 2005 \\
\end{flushright}\end{normalsize}

\begin{center}
  
\vspace{30pt}
  
{\Large \bf Nonlocal Effective Field Equations for Quantum Cosmology }

\vspace{40pt}

{\sl H. W. Hamber}
$^{}$\footnote{e-mail address : HHamber@uci.edu} \\
Department of Physics and Astronomy \\
University of California \\
Irvine, CA 92697-4575, USA \\

\vspace{10pt}

and

\vspace{10pt}

{\sl R. M. Williams}
$^{}$\footnote{e-mail address : R.M.Williams@damtp.cam.ac.uk} \\
Girton College, Cambridge CB3 0JG, and   \\
Department of Applied Mathematics and Theoretical Physics \\
Centre for Mathematical Sciences \\
Wilberforce Road, Cambridge CB3 0WA, United Kingdom.
\\

\end{center}

\vspace{30pt}

\begin{center} {\bf ABSTRACT } \end{center}

\noindent

The possibility that the strength of gravitational
interactions might slowly increase with distance, is explored by formulating
a set of effective field equations, which
incorporate the gravitational, 
vacuum-polarization induced, running of Newton's constant $G$.
The resulting long distance
(or large time) behaviour depends on only one
adjustable parameter $\xi$, and the implications for the 
Robertson-Walker universe are calculated, predicting an accelerated
power-law expansion at later times $t \sim \xi \sim 1/H$.

\vfill


\pagestyle{empty}

\newpage

\pagestyle{plain}

\vskip 10pt
\hspace*{\parindent}

In the Standard Model of particle interactions,
all gauge couplings are known to run with energy. Recent
non-perturbative studies of quantum gravity have suggested that the
gravitational coupling too may depend on a scale related to curvature,
and therefore macroscopic in size. In this Letter, we investigate the
effects of a running gravitational constant $G$ at large
distances. This scale dependence is assumed to be driven by gravitational
vacuum polarization effects, which produce an anti-screening effect
some distance away from the primary source, and therefore tend
to increase the strength of the gravitational coupling.
A power law running of $G$ will
be implemented via manifestly covariant nonlocal terms in the effective
gravitational action and field equations.

We start by assuming ~\cite{det,critical} that for Newton's constant one has
\beq
G(r) \; = \; G(0) \left [ \; 1 \, + \, c_\xi \, ( r / \xi )^{1 / \nu} \, 
+ \, O (( r / \xi )^{2 / \nu} ) \; \right ] \; \; ,
\label{eq:grun}
\eeq
where the exponent $\nu$ is generally related to the derivative
of the beta function for pure gravity evaluated at the non-trivial ultraviolet
fixed point. Recent studies have $\nu^{-1}$ varying between 3.0  and 1.7
~\cite{critical,epsilon,epsilon1,litim,reuter}. 
These estimates rely on three different, and unrelated, nonperturbative
approaches to quantum gravity, based on the lattice path integral formulation,
the two plus epsilon expansion of continuum gravity, 
and a momentum slicing scheme
combined with renormalization group methods in the vicinity of flat space,
respectively.
In all three approaches a non-vanishing, positive bare cosmological constant
is required for the consistency of the renormalization group procedure.
The mass scale $m = \xi^{-1}$ in Eq.~(\ref{eq:grun}) is supposed to
determine the magnitude of quantum deviations from the classical theory.
It seems natural to identify  $ 1 / \xi^2 $ with either some 
very large average spatial curvature scale,
or perhaps more appropriately with the Hubble 
constant (as measured today) determining the macroscopic expansion rate of the
universe, via the correspondence
\beq
\xi \; = \; 1 / H \;\; ,
\label{eq:hub}
\eeq
in a system of units for which the speed of light equals one.
A possible concrete scenario is one in which 
$\xi^{-1} = H_\infty = \lim_{t \rightarrow \infty} H(t) =
\sqrt{\Omega_\Lambda}\, H_0$
with $H_\infty^2 = { \Lambda \over 3 } $,
where $\Lambda$ is the observed cosmological constant, and for which
the horizon radius is $ H_{\infty}^{-1}$.

As it stands, the formula for the running of $G$ is coordinate
dependent, and we therefore replace it with a manifestly covariant
expression involving the covariant d'Alembertian operator
\beq
\Box \; = \; g^{\mu\nu} \, \nabla_\mu \nabla_\nu \; \; ,
\eeq
\noindent whose Green's function in $d$ spatial dimensions is known to
behave as
\beq
< x | \, { 1 \over \Box } \, | y > \delta ( r - d(x,y \, \vert \,g) )
\; \sim \; { 1 \over r^{D-2}} \; \; ,
\eeq
\noindent where $d$ is the minimum distance between points $x$ and $y$
in a background with metric $g_{\mu\nu}$. We therefore write, in four
dimensions, 
\beq
G \; \; \rightarrow \; \; G( \Box ) \; = \; G(0) \left [ \; 1 \, + \,
c_\Box \, \left( { 1 \over {\xi^2 \Box}} \right)^{1 / {2\nu}} \, + \, O ((\xi^2 \Box )^{-
1 / \nu} ) \; \right ] \; \; .
\eeq

One way of incorporating this is to replace the gravitational action
\beq
I \; = \; { 1 \over 16 \pi \, G } \int dx \sqrt{g} \, R
\eeq
\noindent by
\beq
I  \; = \; { 1 \over 16 \pi \, G} \int dx \sqrt{g} \,
\left( 1 \, - \, c_{\Box} \, 
\left ( { 1 \over \xi^2 \Box } \right )^{ 1 / 2 \nu} \, 
+ \, O (( \xi^2 \Box )^{- 1 / \nu} ) \; \right ) R \; \; .
\label{eq:ieff_r}
\eeq
The above prescription has in fact been used successfully 
to systematically incorporate the effects of radiative corrections
in an effective action formalism \cite{vilko,bmv}.
It should be noted that
the coefficient $c_\xi$ in Eq.~(\ref{eq:grun})
is expected to be a calculable
number of order one, not necessarily the same
as the coefficient $c_\Box$, as $r$ and $1 / \sqrt{\Box}$ are clearly
rather different entities to begin with.
One should recall here that in general the form of the covariant
d'Alembertian operator $\Box$
depends on the specific tensor nature of the object it is acting on.

The details of the incorporation of this modified $G$ in the
gravitational side of the Einstein equations are given elsewhere
~\cite{eff}. Here
we shall describe instead its incorporation on the matter side of
Einstein's equations, giving the effective field equations 
\beq
R_{\mu\nu} \, - \, \half \, g_{\mu\nu} \, R \, + \, \Lambda \, g_{\mu\nu}
\; = \; 8 \pi G  \, \left( 1 + A( \Box ) \right) \, T_{\mu\nu} \; \; ,
\label{eq:naive_t}
\eeq
\noindent where we have replaced $G(r)$ by $G(0)(1+A(\Box))$. These can
be written in the form  
\beq
R_{\mu\nu} \, - \, \half \, g_{\mu\nu} \, R \, + \, \Lambda \, g_{\mu\nu}
\; = \; 8 \pi G  \, {\tilde T_{\mu\nu}} \; \; ,
\eeq
with $ {\tilde T_{\mu\nu}} \, = \, \left( 1 + A( \Box ) \right) \, T_{\mu\nu}$
defined as an effective, or gravitationally dressed, energy-momentum tensor.
Just like the ordinary Einstein gravity case,
in general ${\tilde T_{\mu\nu}}$ might not be covariantly conserved a priori,
$\nabla^\mu \, {\tilde T_{\mu\nu}} \, \neq \, 0 $, but ultimately the
consistency of the effective field equations demands that it
be exactly conserved in consideration of the Bianchi identity satisfied
by the Riemann tensor.
The ensuing new covariant conservation law
\beq
\nabla^\mu \, {\tilde T_{\mu\nu}} \; \equiv \; 
\nabla^\mu \, \left [ \left ( 1 + A( \Box ) \right ) \, T_{\mu\nu} 
\right ] \;  = \;  0
\label{eq:continuity}
\eeq
can be then be viewed as a constraint
on ${\tilde T_{\mu\nu}}$ (or $T_{\mu\nu}$) which, for example,
in the specific case of a perfect fluid, 
will imply again a definite relationship between the density $\rho(t) $,
the pressure $p(t)$ and the Robertson-Walker scale factor $R(t)$, just
as it does in the standard case.

For simplicity we set the cosmological constant $\Lambda$ to zero from
now on and consider first the trace of the effective field equations 
\beq
R \; = \; 8 \pi G  \, \left( 1 + A( \Box ) \right) \,
T_{\mu}^{\;\;\mu} \; \; .
\label{eq:naive_tt}
\eeq
The advantage of this is that, initially, we need to consider the
action of $\Box$ only on a scalar function, $S(x)$ say, which is given
by 
\beq
{1 \over \sqrt{g} } \, \partial_\mu \, g^{\mu\nu} \sqrt{g} \,
\partial_\nu \, S(x) \; \; .
\eeq
For the Robertson-Walker metric,
\beq
ds^2 \; = \; - dt^2 + R^2(t) \, \left \{ { dr^2 \over 1 - k\,r^2 }
+ r^2 \, \left( d\theta^2 + \sin^2 \theta \, d\varphi^2 \right)
\right \} \; \; ,
\eeq
$\Box$ acting on a scalar function of $t$ only, is  
\beq
- { 1 \over R^3(t) } \, { \partial \over \partial t } \left [ 
R^3(t) { \partial \over \partial t } \right ] \; \; . 
\eeq

The energy-momentum tensor for a perfect fluid is given by
\beq
T_{\mu\nu} \; = \; \left( p(t) \, + \, \rho(t) \right) u_\mu \, u_\nu
\, + \, g_{\mu\nu} \, p(t) \; \; .
\label{eq:perfect}
\eeq
We consider a pressureless fluid $p(t)=0$ and assume the density and
scale factor are given by powers of $t$, as in the classical solution
for the RW metric: $\rho(t)=\rho_0 t^{\beta}, R(t)=R_0
t^{\alpha}$. Then 
\beq
\Box^n \left( T_{\mu}^{\;\;\mu} \right) \; = \; \Box^n \left( - \rho(t) \right) \; \rightarrow \; 4^n (-1)^{n+1}  
{ \Gamma ( {\beta \over 2} + 1 ) 
  \Gamma ( { \beta + 3 \alpha + 1 \over 2 } ) 
\over
  \Gamma ( {\beta \over 2} + 1 - n ) 
  \Gamma ( { \beta + 3 \alpha + 1 \over 2 } - n  )  } \; \rho_0 \,
  t^{\beta - 2n} \; \; .
\eeq
We may analytically continue the exponent to negative fractional $n$,
~\cite{frac} and obtain with $n =-1/(2\nu)$, an expression for $( 1 +
A(\Box))$ acting on the trace of $T_{\mu\nu}$, given by
\beq
- \left( 1 + c_{\xi} \, \left( { t \over \xi } \right)^{1/\nu} \,
\right) \, \rho_0 \, t^{\beta} \; \; ,
\eeq
with 
\beq
c_{\nu} \, = \, 4^{-1/{2\nu}} \, (-1)^{1 - {1/{2\nu}}}
{ \Gamma ( {\beta \over 2} + 1 )
  \Gamma ( {\beta + 3 \alpha + 1 \over 2 } )
\over
  \Gamma ( {\beta \over 2} + 1 + { 1 \over 2 \nu} )
  \Gamma ( {\beta + 3 \alpha + 1 \over 2 } + { 1 \over 2\nu} )} \; \; .
\eeq
Using the value of the scalar curvature for the Robertson-Walker
metric in the $k=0$ case,
\beq
R \; = \; 6 \left( {\dot{R}}^2(t) + R(t)\,\ddot{R}(t) \right) / R^2(t)
\; \; ,
\label{eq;scalar}
\eeq
gives 
\beq
{ 6 \, \alpha \, \left( 2\alpha - 1 \right) \over t^2 } \, = \, -
\left( 1 + c_{\xi} \, \left( { t \over \xi } \right)^{1/\nu} \right) \,
\rho_0 \, t^{\beta} \; \; .
\eeq
For large $t$, when the correction term starts to take over, we see
from the powers of $t$ that 
\beq
\beta \; = \; - 2 - {1/\nu} \; \; .
\label{eq:beta}
\eeq

Next we will examine the full effective field equations (as opposed
to just their trace part) of Eq.~(\ref{eq:naive_t}) with $\Lambda=0$,
\beq
R_{\mu\nu} \, - \, \half \, g_{\mu\nu} \, R \, 
\; = \; 8 \pi G  \, \left( 1 + A( \Box ) \right) \, T_{\mu\nu} \; \; .
\eeq
Here the d'Alembertian operator
\beq
\Box \; = \; g^{\mu\nu} \, \nabla_\mu \nabla_\nu 
\eeq
acts on a second rank tensor,
\bea
\nabla_{\nu} T_{\alpha\beta} \, = \, \partial_\nu T_{\alpha\beta} 
- \Gamma_{\alpha\nu}^{\lambda} T_{\lambda\beta} 
- \Gamma_{\beta\nu}^{\lambda} T_{\alpha\lambda} \, \equiv \, I_{\nu\alpha\beta}
\nonumber
\eea
\beq 
\nabla_{\mu} \left( \nabla_{\nu} T_{\alpha\beta} \right)
= \, \partial_\mu I_{\nu\alpha\beta} 
- \Gamma_{\nu\mu}^{\lambda} I_{\lambda\alpha\beta} 
- \Gamma_{\alpha\mu}^{\lambda} I_{\nu\lambda\beta} 
- \Gamma_{\beta\mu}^{\lambda} I_{\nu\alpha\lambda} \; \; ,
\eeq
and would thus seem to require the calculation of 1920 terms,
of which fortunately many vanish by symmetry.
Assuming that $T_{\mu\nu}$ describes a pressureless perfect fluid, we obtain
 
\bea
\left( \Box \, T_{\mu\nu} \right )_{tt} \; & = & \; 
6 \, \rho (t) \,
\, \left ( { \dot{R}(t) \over R(t) } \right )^2
\, - \, 3 \, \dot{\rho}(t) \,  { \dot{R}(t) \over R(t) }
\, - \, \ddot{\rho}(t) 
\nonumber \\
\left( \Box \, T_{\mu\nu} \right )_{rr} \; & = & \; 
{ 1 \over 1 \, - \, k \, r^2 } \left(
2 \, \rho (t) \, \dot{R}(t)^2 
\, \right)
\nonumber \\
\left( \Box \, T_{\mu\nu} \right )_{\theta\theta} \; & = & \; 
r^2 \, ( 1 \, - \, k \, r^2 ) \, 
\left( \Box \, T_{\mu\nu} \right )_{rr}
\nonumber \\
\left( \Box \, T_{\mu\nu} \right )_{\varphi\varphi} \; & = & \; 
r^2 \, ( 1 \, - \, k \, r^2 ) \, 
\sin^2 \theta \, \left( \Box \, T_{\mu\nu} \right )_{rr} \; \; ,
\label{eq:boxont}
\eea
with the remaining components equal to zero.
Note that a non-vanishing pressure contribution is generated in the effective
field equations, even if one assumes initially a pressureless fluid.
As before, repeated applications of the d'Alembertian $\Box$ to the above expressions leads
to rapidly escalating complexity (for example, eighteen distinct 
terms are generated by $\Box^2$ for each of the above contributions),
which can only be tamed by introducing some further simplifying assumptions.
In the following we will therefore assume as before that $k=0$,
$\rho(t) = \rho_0 \, t^\beta$, and $R(t) = R_0 \, t^\alpha $. We obtain
\bea
\left( \Box \, T_{\mu\nu} \right )_{tt} \; & = & \; 
\left( 6 \, \alpha^2 - \beta^2 - 3 \, \alpha \, \beta  + \beta \right) \, \rho_0 \,
t^{\beta - 2 }
\nonumber \\
\left( \Box \, T_{\mu\nu} \right )_{rr} \; & = & \; 
2 R_0^2 \, t^{2 \alpha} \alpha^2  \, \rho_0 \, t^{\beta -2 } \; \; ,
\eea
which again shows that the $tt$ and $rr$ components get mixed by the
action of the $\Box$ operator, and that a 
non-vanishing $rr$ component gets generated,
even though it was not originally present.

The geometric side of the gravitational field equations, the Einstein
tensor, has the following components for the RW metric:
\bea
G_{tt} \; & = & \; 3 \, {\dot{R}}^2 (t) / R^2(t)
\nonumber \\
G_{rr} \; & = & \; { -1 \over 1 - k\, r^2 }  \left( \dot{R}^2(t) + 2 \,
R(t) \, \ddot{R}(t) \right)
\nonumber \\
G_{\theta\theta}\; & = & \; r^2 \left( 1 - k \, r^2 \right) G_{rr}
\nonumber \\
G_{\varphi\varphi} \; & = & \; \sin^2 {\theta} \, G_{\theta\theta} \; \; .
\eea
Then with $k=0$ and $R(t) = R_0 \, t^{\alpha}$, these will all behave
like $t^{-2}$ so in fact a solution can only be achieved at order $\Box^n$ 
provided the exponent $\beta$ satisfies $\beta = -2 + 2 n $, or
since $n=-1/( 2 \nu )$,
\beq
\beta \; = \; - 2 \, - \, 1 / \nu \; \; ,
\label{eq:beta1}
\eeq
as was found previously from the trace equation,
Eqs.~(\ref{eq:naive_tt}) and (\ref{eq:beta}).

We must now determine $\alpha$. By repeated application of $\Box$,
for general $n$ one can then write 
\beq
\left( \Box^n \, T_{\mu\nu} \right )_{tt} \; \rightarrow \; 
c_{tt} ( \alpha, \nu ) \, \rho_0 \, t^{- 2 } 
\eeq
and similarly for the $rr$ component
\beq
\left( \Box^n \, T_{\mu\nu} \right )_{rr} \; \rightarrow \; 
c_{rr} ( \alpha, \nu ) \, R_0^2 \, t^{2 \alpha} \, \rho_0 \, t^{- 2 }
\; \; . 
\eeq
But remarkably (see also Eq.~(\ref{eq:boxont}) )
one finds for the two coefficients the simple identity
\beq
c_{rr} (\alpha, \nu ) \; = \; \third \, c_{tt} ( \alpha, \nu)
\label{eq:crr}
\eeq
as well as $ c_{\theta\theta} = r^2 \, c_{rr} $ and
$ c_{\varphi\varphi} = r^2 \, \sin^2 \theta \, c_{rr} $.
Then for large times, when the quantum correction starts to become important,
the $tt$ and $rr$ field equations reduce to
\beq
3 \, \alpha^2 \, t^{-2} \; = \; 8 \pi G \,
c_{tt} ( \alpha, \nu ) \, \rho_0 \, t^{- 2 } 
\eeq
and 
\beq
- \; \alpha \, ( 3 \, \alpha \, - \, 2 ) \, R_0^2 \, t^{2 \alpha - 2 } 
\; = \; 8 \pi G \,
c_{rr} ( \alpha, \nu ) \, R_0^2 \, t^{2 \alpha} \, \rho_0 \, t^{- 2 } 
\eeq
respectively.
But the identity $ c_{rr} \; = \; \third \, c_{tt} $
implies, simply from the consistency of the $tt$ and $rr$ effective
field equations at large times,
\beq
{ c_{rr} ( \alpha, \nu)  \over c_{tt} (\alpha, \nu ) } \; \equiv \; \third 
\; = \; - \, { 3 \alpha - 2  \over 3 \alpha } \; \; ,
\eeq
whose only possible solution finally gives the second sought-for result, namely
\beq
\alpha \; = \; { 1 \over 2 } \; \; .
\label{eq:alpha}
\eeq

We still need to check whether the above solution is consistent
with covariant energy conservation.
With the assumed form for $T_{\mu\nu}$
it is easy to check that energy conservation yields for the $t$ component
\beq
\left ( \nabla^\mu \, \left ( \Box^n \, T_{\mu\nu} \right ) \right )_t 
\; \rightarrow \; - \, \left ( 
( 3 \, \alpha \, + \, \beta \, + \, 1 / \nu ) \, c_{tt} \, + \,
3 \, \alpha \, c_{rr} 
\right ) \, \rho_0 \, t^{\beta + 1 / \nu - 1} \; = \; 0
\eeq
when evaluated for $n=-1 / 2 \nu$, and zero for the remaining three spatial
components.
But from the solution for the matter density $\rho(t)$ at large times one
has $\beta = - 2 - 1 / \nu $, so the above zero condition gives again
$ c_{rr} / c_{tt} = - ( 3 \alpha - 2 )/ 3 \alpha $, exactly 
the same relationship previously implied by the consistency of the
$tt$ and $rr$ field equations.

Let us emphasize that the values for $\alpha= 1/2 $ of Eq.~(\ref{eq:alpha}) and 
$\beta =-2 -1/ \nu $ of Eq.~(\ref{eq:beta1}), determined from
the effective field equations at large times, are found to be consistent
with {\it both} the field equations {\it and} covariant energy conservation.
More importantly, the above solution is also consistent with what was found
previously by looking at the trace of the effective field equations,
Eq.~(\ref{eq:naive_tt}), which also implied the result
$\beta=-2 -1/ \nu$, Eq.~(\ref{eq:beta}).

The classical unmodified matter-dominated RW equations have solutions
$\alpha=2/3$, $\beta=-2$, which mean that the scale factor behaves as
\beq
R(t) \; \sim \; t^{\alpha} \; \sim \; t^{2/3}
\eeq
and the density as
\beq
\rho(t) \; \sim \; t^{\beta} \; \sim \; t^{-2} \; \sim \; \left( (
R(t) \right) ^{-3} \; \; .
\eeq
This will also be the behaviour for our model at early times, but at
later times, when the effect of the gravitational vacuum-polarization
modification dominates, the behaviour is rather different: for the
scale factor, we have
\beq
R(t) \; \sim \; t^{\alpha} \; \sim \; t^{1/2}
\eeq
and for the density
\beq
\rho (t) \; \sim \; t^{\beta} \; \sim \;
t^{- 2 - 1 / \nu } \; \sim \; 
\left ( R(t) \right )^{- 2 \, ( 2 + 1 / \nu ) } \; \; .
\eeq 
Thus the density decreases significantly faster in time
than the classical value ($ \rho (t) \sim t^{-2}$),
a signature of an accelerating expansion at later times.

Within the Friedmann-Robertson-Walker (FRW) framework
the gravitational vacuum polarization term behaves in many ways
like a positive pressure term.
The value $\alpha=1/2$ corresponds to $\omega=1/3$ in
\beq
\alpha \; = \; { 2 \over 3 ( 1 \, + \, \omega ) } \; \; ,
\eeq
(this follows from the consistency of the $rr$ and $tt$ equations in
the general case) where we have taken the pressure and density to be related by 
$p(t) = \omega \, \rho(t)$, which is therefore characteristic of radiation.
One can therefore visualize
the gravitational vacuum polarization contribution
as behaving like ordinary radiation, in the form of a dilute virtual
graviton gas.
It should be emphasized though that the relationship between density $\rho (t)$
and scale factor $R(t)$ is very different from the classical case.

\vspace{20pt}

{\bf Acknowledgements}

The authors are grateful to Gabriele Veneziano for his close involvement
in the early stages of this work, and for bringing
to our attention the work of Vilkovisky and collaborators
on nonlocal effective actions for gauge theories. 
Both authors also wish to thank the Theory Division at CERN
for warm hospitality and generous financial support during the
Summer of 2004.
One of the authors (HWH) wishes to thank James Bjorken for
useful discussions on issues related to the subject
of this paper.
The work of Ruth Williams was supported in part by the UK Particle
Physics and Astronomy Research Council.


\vfill

\newpage

\end{document}